\documentclass[aps,prb,twocolumn,showpacs,preprintnumbers,
amsmath,amssymb,superscriptaddress]{revtex4}
\usepackage{graphicx}
\usepackage{bm}

\begin{document}

\title{Proximity-effect-induced superconductivity in topological insulator-related material Bi$_2$Se$_3$}

\author{Fan Yang}
\affiliation{Daniel Chee Tsui Laboratory, Beijing National
Laboratory for Condensed Matter Physics, Institute of Physics,
Chinese Academy of Sciences, Beijing 100190, China}
\author{Fanming Qu}
\affiliation{Daniel Chee Tsui Laboratory, Beijing National
Laboratory for Condensed Matter Physics, Institute of Physics,
Chinese Academy of Sciences, Beijing 100190, China}
\author{Jie Shen}
\affiliation{Daniel Chee Tsui Laboratory, Beijing National
Laboratory for Condensed Matter Physics, Institute of Physics,
Chinese Academy of Sciences, Beijing 100190, China}
\author{Yue Ding}
\affiliation{Daniel Chee Tsui Laboratory, Beijing National
Laboratory for Condensed Matter Physics, Institute of Physics,
Chinese Academy of Sciences, Beijing 100190, China}
\author{Jun Chen}
\affiliation{Daniel Chee Tsui Laboratory, Beijing National
Laboratory for Condensed Matter Physics, Institute of Physics,
Chinese Academy of Sciences, Beijing 100190, China}
\author{Zhongqing Ji}
\affiliation{Daniel Chee Tsui Laboratory, Beijing National
Laboratory for Condensed Matter Physics, Institute of Physics,
Chinese Academy of Sciences, Beijing 100190, China}
\author{Guangtong Liu}
\affiliation{Daniel Chee Tsui Laboratory, Beijing National
Laboratory for Condensed Matter Physics, Institute of Physics,
Chinese Academy of Sciences, Beijing 100190, China}
\author{Jie Fan}
\affiliation{Daniel Chee Tsui Laboratory, Beijing National
Laboratory for Condensed Matter Physics, Institute of Physics,
Chinese Academy of Sciences, Beijing 100190, China}
\author{Changli Yang}
\affiliation{Daniel Chee Tsui Laboratory, Beijing National
Laboratory for Condensed Matter Physics, Institute of Physics,
Chinese Academy of Sciences, Beijing 100190, China}
\author{Liang Fu}
\affiliation{Department of Physics, Massachusetts Institute of
Technology, Cambridge, MA 02139, U. S. A. }
\author{Li Lu}\email[Corresponding authors: ]{lilu@iphy.ac.cn}
\affiliation{Daniel Chee Tsui Laboratory, Beijing National
Laboratory for Condensed Matter Physics, Institute of Physics,
Chinese Academy of Sciences, Beijing 100190, China}

\date{\today}

\begin{abstract}
We have studied the electron transport properties of topological
insulator-related material Bi$_{2}$Se$_{3}$ near the superconducting
Pb-Bi2Se3 interface, and found that a superconducting state is
induced over an extended volume in Bi$_{2}$Se$_{3}$. This state can
carry a Josephson supercurrent, and demonstrates a gap-like
structure in the conductance spectra as probed by a normal-metal
electrode. The establishment of the gap is not by confining the
electrons into a narrow space close to the superconductor-normal
metal interface, as previously observed in other systems, but
presumably via electron-electron attractive interaction in
Bi$_{2}$Se$_{3}$.
\end{abstract}



\maketitle

Through proximity effect (PE) between a superconductor and a normal
metal \cite{PE_review}, pairing correlation between electrons is
delivered from the superconducting side to the normal-metal side. PE
has been intensely studied over several decades. It is believed that
a PE-affected normal metal would remain resistive unless an energy
gap is established to prevent the electrons from scattering
\cite{non-superconducting01, non-superconducting02}. According to
McMillan \cite{McMillan}, this energy gap could be established
either through electron-electron (e-e) attractive interaction, or by
confining the electrons into a narrow space close to the S-N
interface (i.e., in a confined S-N structure), where S and N denote
superconductor and normal metal, respectively. From the experimental
side, the existence of a PE-induced gap
\cite{PE_gap01,PE_gap02,PE_gap03,PE_gap04,JiaJinFeng} and the
ability of carrying a supercurrent \cite{PE_gap04} have long been
verified in confined S-N structures. However, in open S-N structures
where the electrons in the normal metal are not necessarily confined
near the interface, neither the existence of a PE-induced
supercurrent-carrying state nor the existence of an energy gap has
been verified experimentally by using independent probes/contacts,
only a gap-like structure (i.e., not a true superconducting gap) was
observed via a tunneling measurement \cite{PE_gap05}.

Recently, particular attention has been paid to the proximity effect
between an $s$-wave superconductor and a topological insulator (TI)
\cite{TI1,TI2,TI3}. An unconventional superconducting state
resembling a spinless $p_{x}+ip_{y}$ superconductor is expected to
occur at the S-TI interface. And the vortex cores of that state are
believed to host Majorana fermions based on which topological
quantum computation can be realized \cite{MF1,MF2,MF3}. To explore
these novel phenomena experimentally, one would first need to
identify the existence of a PE-induced superconducting state at the
S-TI interface. So far, a supercurrent has been observed in S-TI-S
type of devices such as Al-Bi$_2$Se$_3$-Al \cite{Al-Bi2Se3},
W-Bi$_2$Se$_3$-W \cite{W-Bi2Se3}, Pb-Bi$_2$Te$_3$-Pb
\cite{Pb-Bi2Te3}, and Nb-Bi$_2$Te$_3$ \cite{Nb-Bi2Te3}. However,
since a supercurrent in S-N-S devices can be carried not only by a
PE-induced superconducting state, but also by phase-dependent
Andreev bound states \cite{PE_review, SNS01, SNS02}, with these
results we cannot identify the existence of an independent
superconducting state in the PE-affected region. Recently, a
superconducting energy gap was observed in ultra-thin Bi$_2$Se$_3$
films grown on a superconducting NbSe$_2$ substrate
\cite{JiaJinFeng}. It proves that a PE-induced superconducting state
can be formed in Bi$_2$Se$_3$ films when the electrons there are
confined to be close to the S-TI interface, similar to those
reported in the early literature
\cite{PE_gap01,PE_gap02,PE_gap03,PE_gap04}. But it would still be
interesting to investigate whether a superconducting state can be
established in the TI side without the structural confinement to the
electrons. And if yes, how far this state could survive away from
the S-TI interface.

Here, we report the observation of a PE-induced zero-resistance
state in Bi$_2$Se$_3$ over an extended distance of $\sim 1\mu$m away
from the Pb-Bi$_{2}$Se$_{3}$ interface. The critical supercurrent of
the zero-resistance state exhibits a Fraunhofer diffraction pattern
against magnetic field. Moreover, a gap-like structure has also been
observed in the PE-affected region. The results suggest that the
PE-affected region can be regarded as a weak and independent
superconductor.

Bi$_{2}$Se$_{3}$ flakes were exfoliated onto SiO$_2$-Si substrates
from a high-quality single crystal. Those with thickness of
$\sim$100 nm were selected out and further fabricated into devices
using a standard e-beam lithography technique. All of the metal
films were deposited via magnetron sputtering. The Pb film were
deposited lastly, to avoid otherwise being baked at 180$^{\circ}$C
during the fabrication of other electrodes, a process which might
alloy the Pb-Bi$_{2}$Se$_{3}$ interface \cite{Pb-Bi2Te3}.

Figure 1 (a) and (b) show the scanning electron microscope (SEM)
images of a typical device used in this experiments. A
superconducting Pb film was deposited at the center of the
Bi$_2$Se$_3$ flake, and two large Pd electrodes were deposited at
each end of the flake. Between the Pb film and each of the large Pd
electrodes, two small Pd electrodes were deposited, one located
close to the Pb film ($\sim$100 nm away) and the other about 1
$\mu$m away. These Pd electrodes allow us to perform four-terminal
resistance measurement in the PE-affected area on the Bi$_2$Se$_3$
flake, and also to probe the local electronic density of states
(DOS) there. We have investigated two devices of this type and
obtained very similar results. Investigations on three more devices
with a slightly different alignment of Pd electrodes also gave a
consistent conclusion. In the following we present the data taken
from the device shown in Fig. 1 (a).

\begin{figure}
\includegraphics[width=1 \linewidth]{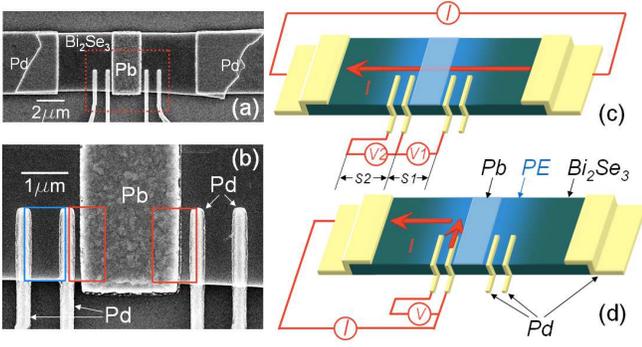}
\caption{\label{fig:Fig1} {(color online) (a) and (b) SEM images of
the device. The red dotted rectangle area in (a) is magnified in
(b). The red and blue rectangles in (b) illustrate the possible
areas responsible for the periods of the Fraunhofer diffraction
patterns in Figs. 2(a) and (b), respectively. (c) Illustration of
the measurement configuration for probing the resistance of the
PE-affected region. The sections measured by lock-in amplifiers V1
and V2 are denoted as S1 and S2, respectively. The areas in
light-blue near the Pb film indicate the PE-affected regions. (d)
Illustration of a three-terminal measurement configuration for
probing the conductance spectrum of a Pd-Bi$_2$Se$_3$ contact in the
PE-affected region.}}
\end{figure}

\begin{figure}
\includegraphics[width=0.8 \linewidth]{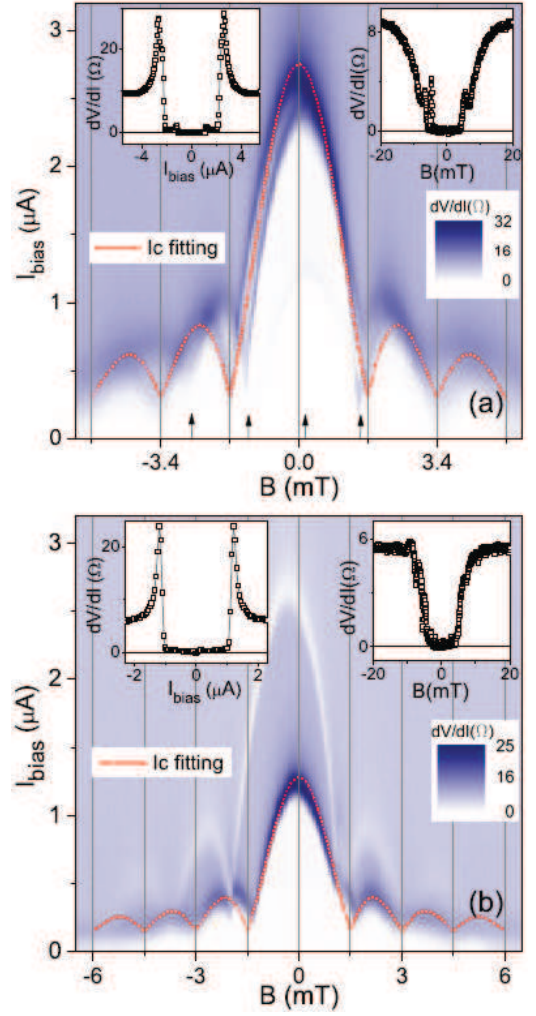}
\caption{\label{fig:Fig2} {(color online) (a) 2D plot of the $dV/dI$
data for S1 measured as a function of magnetic field and dc bias
current. The white area represents a zero resistance state. The grid
lines and the arrows show the minima of the two superimposed
Fraunhofer diffraction patterns. The possible effective areas
corresponding to the two periods of the patterns are indicated in
Fig. 1(b) by the red rectangles. The red dotted line is a fitting
curve discussed in the text. Left inset: $dV/dI$ versus bias current
in zero field. The two down arrows indicate the echo of the
superconducting transition of S2. Right inset: zero-bias $dV/dI$
versus magnetic field. (b) 2D plot of the $dV/dI$ data for S2.
Besides the main Fraunhofer pattern, echo from the superconducting
transition of S1 can be clearly seen as a thin-white curve. Left
inset: $dV/dI$ versus bias current in zero field. Right inset:
zero-bias $dV/dI$ versus magnetic field.}}
\end{figure}

Two different measurement configurations were employed, as
respectively illustrated in Figs. 1(c) and (d). In order to measure
the four-terminal resistance of the PE-affected region, we pass a
current through Bi$_2$Se$_3$ crystal from the large Pd electrodes at
the two ends. Then the resistance of sections S1 and S2 was measured
via corresponding Pd electrodes. While probing the electronic DOS of
the PE-affected region, the contact resistance of a small Pd
electrode near the Pb film was measured by using a three-terminal
configuration. All measurements were carried out in a top-loading
dilution refrigerator with a base temperature of 15 mK. Lock-in
amplifiers were used, with an ac excitation current of 50 nA and at
30.9 Hz. A Keithley 2400 source-meter was used to drive the
superconducting magnet, to guarantee a precise control of magnetic
field at milli-Gauss level.

In Figs. 2 (a) and (b) we show the 2D mappings of $dV/dI$ data as a
function of both magnetic field and dc bias current for sections S1
and S2, respectively. The zero magnetic field is determined
according to the symmetry of the main structure of the data, which
is slightly shifted for different sections, presumably due to local
flux pining in the Pb film. A zero-resistance superconducting state
was found at low temperatures, represented by the white areas. This
state can be destructed by increasing bias current and/or magnetic
field, as shown in the insets of Figs. 2(a) and (b). The critical
supercurrents vary with applied magnetic field, following a
Fraunhofer diffraction pattern. In Fig. 2(a), we can see two sets of
Fraunhofer patterns superimposed onto each other, with slightly
different periods, as evident in the positions of the pattern minima
indicated by the grid lines and the arrows, respectively. We would
tentatively ascribe these two sets of patterns to the two S1
sections with different areas at the two sides of the Pb film.

The Fraunhofer pattern shown in Fig. 2(b) was measured across S2. In
this region, the induced superconducting state is not as strong as
in S1 because of its farther distance from the Pb-Bi$_{2}$Se$_{3}$
interface. Therefore, its critical supercurrent is smaller. Because
the whole device is at a mesocopic scale, a strong echo from the
superconducting transition of S1 can be seen as a thin-white curve
in the 2D plot for S2. Similarly, a faint echo from the transition
of S2 can also be recognized in Fig 2(a), and is best seen in the
left inset of Fig. 2(a) where the conductance undergoes small but
sudden increases at $I_{\rm{bias}}\simeq\pm 1.3 \mu$A, as marked by
the down arrows.

The data can be phenomenologically fitted to a Fraunhofer-like
diffraction pattern:
$$I_{\rm c}(B)=I_{\rm c0} \left|\sin\left(\frac{ \pi\phi_{\rm J}}
{\phi_{\rm 0}}\right)/\left(\frac{\pi\phi_{\rm J}}{\phi_{\rm 0}}
\right)\right|+I_{\rm c1}$$ where $\phi_{\rm J}$ is the magnetic
flux through some effective area, $\phi_{\rm 0}=h/(2e)$ is the flux
quanta, and $I_{\rm c1}$ represents the contribution from a part of
PE-induced superconducting area which is not included in the
effective area causing the oscillation, but is shunted to it. The
non-uniform distribution of the PE-induced superconductivity
\cite{Pb-Bi2Te3}, hence the supercurrent density, might also play a
role. The fitting curves are shown in Figs. 2(a) and (b) as red
dotted lines. We have $I_{\rm c0}$=2.45 $\mu$A and $I_{\rm c1}$=0.3
$\mu$A for the fitting in Fig. 2(a), and $I_{\rm c0}$=1.13 $\mu$A
and $I_{\rm c1}$=0.15 $\mu$A for the fitting in Fig. 2(b).

From the fittings we can calculate the effective area of the
sections: $S_{\rm eff}=\phi_{\rm 0}/\Delta B$, where $\Delta B$ is
the period of the Fraunhofer-like pattern. For example, the two
periods of 1.7 mT and 1.38 mT shown in Fig. 2(a) correspond to
effective areas of 1.22 $\mu m^{2}$ and 1.50 $\mu m^{2}$,
respectively, as tentatively ascribed to the two red rectangles in
Fig. 1(b). These effective areas appear to be larger than the actual
openings between the Pb film and Pd electrodes because of flux
compressing and penetration at the edge of the superconducting Pb
film. The period of 1.5 mT in Fig. 2(b) corresponds to an effective
area of 1.38 $\mu m^{2}$, as represented by the blue rectangle in
Fig. 1(b). Flux compressing is less pronounced for S2 because this
section is located farther from the Pb film. There are more
evidences (data not shown) demonstrating that there exist multiple
sets of Fraunhofer diffraction patterns, whose periods correspond to
some characteristic areas defined by different pairs of normal-metal
electrodes. The underline mechanism of this phenomenon is, however,
not obvious and warrants further investigations.

\begin{figure}
\includegraphics[width=0.85 \linewidth]{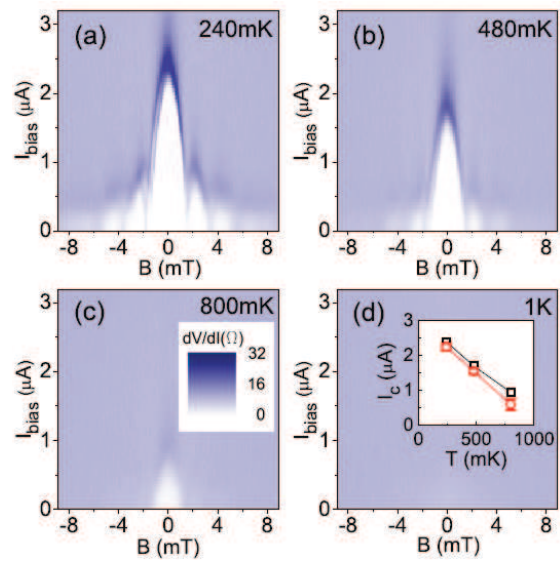}
\caption{\label{fig:Fig3} {(color online) 2D plots of $dV/dI$ of S1
as a function of magnetic field and dc bias current at (a) 240 mK,
(b) 480 mK, (c) 800 mK and (d) 1K. The inset shows the temperature
dependence of the zero-field peak height of the Fraunhofer patterns.
Black squares: using $dV/dI$ maximum as the criteria. Red squares:
using zero resistance as the criteria. }}
\end{figure}

In Fig. 3 we show the 2D plots of the $dV/dI$ data for S1 measured
at different temperatures. The Fraunhofer pattern becomes blurry as
temperature increases, and being featureless at $T$=1 K. The
zero-field peak height decreases linearly with increasing
temperature, as shown in the inset of Fig. 3(d). Similar behavior
was also observed on S2 (data not shown), where the data become
featureless above $T$=800 mK.

\begin{figure}
\includegraphics[width=0.9 \linewidth]{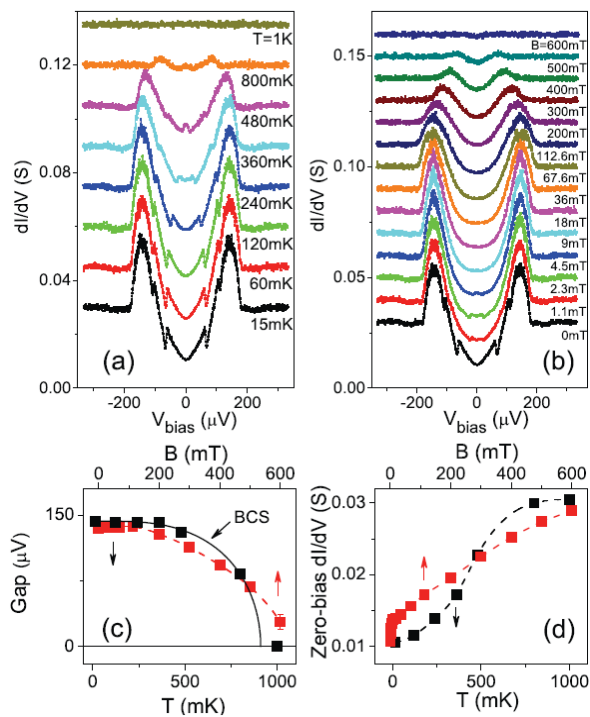}
\caption{\label{fig:Fig4} {(color online) (a) Conductance spectra of
a Pd contact 95 nm away from the Pb film, measured in a
three-terminal configuration shown in Fig. 1(d). The curves other
than the 15 mK one are shifted vertically for clarity. (b)
Conductance spectra of the same contact measured at 15 mK and in
different magnetic fields. Again, the curves other than the one
taken in zero field are shifted vertically for clarity. (c) Peak
position plotted as functions of temperature (black squares) and
magnetic field (red squares). The black line shows the
gap-temperature relation expected from the BCS theory. (d) Zero-bias
conductance at different temperatures (black squares) and in
different fields (red squares).}}
\end{figure}

The observed Fraunhofer-like field dependent envelops unambiguously
imply that the supercurrents between the Pb and Pd electrodes, and
even between two normal-metal Pd electrodes, are all Josephson
supercurrents. It also guarantees that the white areas within the
envelops represent a superconducting state.

As mentioned in the introduction, in order to sustain a supercurrent
in the PE-affected region, there must exist an energy gap; pair
correlation alone is not sufficient
\cite{non-superconducting01,non-superconducting02}. According to the
theory \cite{McMillan}, electrons in a PE-affected normal metal must
either have an attractive interaction or experience a confinement in
order to establish the energy gap. In a recent
experiment\cite{Au-Al}, the synergetic effects of electron
confinement and attractive interaction have been revealed by
scanning tunneling spectrum studies in Au-Al bi-layer structures.
For our device, the Bi$_{2}$Se$_{3}$ crystal is long enough
($\sim$10 $\mu$m) along the lateral direction from the
Pb-Bi$_{2}$Se$_{3}$ interface, thus no obvious confinement to the
electrons exists. We therefore speculate that the e-e attractive
interaction in Bi$_{2}$Se$_{3}$ plays a crucial role for the
establishment of the gap, and hence the superconducting state, in
the PE-affected region.

It has to be noted that, similar to many other experiments on
Bi$_{2}$Se$_{3}$, the crystal used here is not an intrinsic
topological insulator. The Fermi level is shifted into the
conduction band due to disorders (mainly Se vacancies) in the
sample, so that both the surface states and the bulk states could be
involved in/contribute to the proximity effect. Although it is
rather unusual that in this experiment the PE-affected region could
develop to a distance as far as one micron, which might be
attributed to the novelty of the surface states, we are unable to
give a definite evidence for this point. On the contrary, our
previous work \cite{Pb-Bi2Te3} shows that a strong proximity effect
develops along the thickness direction of the Bi$_{2}$Te$_{3}$
crystal, in addition to along the surface directions of the crystal.
Given the fact that Bi$_{2}$Te$_{3}$ is very similar to
Bi$_{2}$Se$_{3}$ in terms of superconducting PE, we conclude that
bulk electrons should play an important role in the observed
proximity effect, regardless the contribution of the surface
electrons.

In order to detect the expected superconducting gap in the
PE-affected region, we measured the conductance spectra of a small
Pd contact 95 nm away from the Pb-Bi$_{2}$Se$_{3}$ interface. A
three-terminal resistance measurement was performed using the
configuration shown in Fig. 1(d). The conductance spectrum, namely
the bias voltage dependence of differential conductance, was taken
at different temperatures and in different magnetic fields. The
results are shown in Figs. 4(a) and (b). The normal-state resistance
of the contact is 33.7 $\Omega$. Although it is not in the tunneling
limit, according to many experimental studies on S-N junctions
\cite{S-N-beyond-tunneling}, the measured conductance spectrum
should partially reflects the DOS of the electrons beneath the
contact.

From Fig. 4(a), it is evident that a gap-like structure develops
below a critical temperature $T_{\rm c}'\approx$1 K, which is much
lower than the superconducting transition temperature $T_{\rm
c}$=7.2 K of Pb. $T_{\rm c}'\approx$1 K is also the onset
temperature of the Fraunhofer patterns shown in Fig. 3, suggesting a
close connection between these two phenomena, behind which is the
formation of a PE-induced superconducting phase. Plotted in Fig.
4(c) is the gap value of this phase as a function of temperature.
The solid black line is a best fit of the data to the BCS theory. It
yields $T_{\rm c}'=$911 mK.

The gap-like structure disappears in magnetic
fields higher than the critical field $H_{\rm c}$ of the
superconducting Pb film, which is about 600 mT for this device [see
Fig. 4(b)].

For curves taken in zero magnetic field, there are some jumps inside
the gap. While we do not know the origin of these jumps, we
speculate that they might arise from some mesoscopic processes with
trajectories threaded by magnetic flux, because they are quickly
suppressed by raising magnetic field to 1.1 mT. It should be noted
that the jumps are not subgap conductance peaks caused by multiple
Andreev reflections, which are usually seen in S-N-S junctions.

For the curve taken at $T=$15 mK and in zero field, the peak
position of the gap-like structure is 144 $\mu$V, being only 9.5\%
of the gap value of Pb. This result confirms our previous result
obtained on Sn-Bi$_{2}$Se$_{3}$ junctions \cite{Sn-Bi2Se3}, where a
small gap-like structure occurs at about 1/3 of the gap energy of
Sn. It seems to be common that a gap-like structure is induced in
the PE-affected regions, with a gap value significantly smaller than
that of the maternal superconductor.

Question remains as whether the observed gap-like structure in the
conductance spectrum represents a true superconducting gap, since a
non-zero subgap conductance (in this experiment, it is about 36\% of
the normal-state value at 15 mK) could be attributed either to
Andreev reflection at the contact \cite{BTK}, or to a gap-like
structure caused by pair correlation in the PE-affected region
\cite{PE_gap05,Usadel01,Usadel02,Usadel03}. Although the appearance
of a supercurrent in our current experiments strongly suggest the
existence of a true energy gap, more conclusive data should be
obtained via future measurements in the tunneling limit. If a
PE-induced gap does exist, Andreev reflection will be greatly
suppressed at sufficiently low temperatures in the tunneling limit,
resulting in a well-defined gap structure and with a near-zero
subgap conductance. Otherwise if the measured gap-like structure is
caused by pair correlation, the subgap conductance will remain to be
non-zero at low temperatures.

In summary, we have demonstrated that an extended volume in
Bi$_{2}$Se$_{3}$, up to $\sim 1 \mu$m away from the
Pb-Bi$_{2}$Se$_{3}$ interface, can be induced to superconducting
through the proximity effect. This superconducting state can carry a
Josephson supercurrent, whose critical value oscillates with
magnetic field in a Fraunhofer diffraction patten. Since there is no
confinement to the electrons in the PE-affected region of
Bi$_{2}$Se$_{3}$ in our devices, the superconducting gap there,
which is needed for stabilizing the induced superconducting state
and as has emerged in the measured conductance spectra, is
presumably established by e-e attractive interaction in
Bi$_{2}$Se$_{3}$.

\begin{acknowledgments}
We would like to thank T. Xiang, S. P. Zhao, Z. Fang, X. Dai, C.
Ren, L. Shan and X. C. Xie for stimulative discussions. This work
was supported by the National Basic Research Program of China from
the MOST under the contract No. 2009CB929101 and 2011CB921702, by
the NSFC under the contract No. 11174340 and 11174357, and by the
Knowledge Innovation Project and the Instrument Developing Project
of CAS.

\end{acknowledgments}

\end{document}